%% file: main.tex
\RequirePackage{amsmath}
\documentclass[runningheads]{llncs}
\usepackage{graphicx} 
\usepackage{amsmath}
\usepackage{ifthen}
\usepackage{adjustbox}
\usepackage{xkeyval}
\usepackage{tikz}
\usepackage{calc}
\usepackage{amssymb}
\usepackage{multirow}
\usepackage{paralist}
\usepackage{comment}
\usepackage{todonotes}
\usepackage[colorlinks = true,
            linkcolor = black,
            urlcolor  = blue,
            citecolor = blue,
            anchorcolor = black]{hyperref}

\def\eg{\emph{e.g. }}
\def\ie{\emph{i.e. }}

\begin{document}

\title{Evaluation of CT Image Synthesis Methods: From Atlas-based Registration to Deep Learning}

\author{Andreas D. Lauritzen\inst{1,3} \and Xenophon Papademetris\inst{1,2} \and Sergei Turovets\inst{4} \and John A. Onofrey\inst{1}}
\institute{Departments of Radiology \& Biomedical Imaging, 
\and Biomedical Engineering, Yale University, New Haven, CT, USA \\
\email{\{xenophon.papademetris, john.onofrey\}@yale.edu}
\and
Department of Computer Science, University of Copenhagen, Denmark \\
\email{al@di.ku.dk}
\and
Neuroinformatics Center, University of Oregon, USA \\ 
\email{sergei@cs.uoregon.edu}
\\
}

\authorrunning{A. Lauritzen et al.}

\titlerunning{Evaluation of CT Image Synthesis Methods}

\maketitle

\begin{abstract}
Computed tomography (CT) is a widely used imaging modality for medical diagnosis and treatment.
In electroencephalography (EEG), CT imaging is necessary for co-registering with magnetic resonance imaging (MRI) and for creating more accurate head models for the brain electrical activity due to better representation of bone anatomy.
Unfortunately, CT imaging exposes patients to potentially harmful sources of ionizing radiation.
Image synthesis methods present a solution for avoiding extra radiation exposure.
In this paper, we perform image synthesis to create a realistic, synthetic CT image from MRI of the same subject, and we present a comparison of different image synthesis techniques.
Using a dataset of 30 paired MRI and CT image volumes, our results compare image synthesis using deep neural network regression, state-of-the-art adversarial deep learning, as well as atlas-based synthesis utilizing image registration.
We also present a novel synthesis method that combines multi-atlas registration as a prior to deep learning algorithms, in which we perform a weighted addition of synthetic CT images, derived from atlases, to the output of a deep neural network to obtain a residual type of learning.
In addition to evaluating the quality of the synthetic CT images, we also demonstrate that image synthesis methods allow for more accurate bone segmentation using the synthetic CT imaging than would otherwise be possible by segmenting the bone in the MRI directly.
\end{abstract}
\section{Introduction}
\input{tex/introduction.tex}
\section{Methods}
\input{tex/method.tex}
\section{Results}
\input{tex/experiments.tex}
\section{Discussion and Conclusion}
\input{tex/discussion.tex}
\input{tex/acknowledgements.tex}
\bibliographystyle{splncs04}
\bibliography{refs}
\setcounter{tocdepth}{1}
\end{document}

%% file: tex/introduction.tex
Magnetic resonance (MR) imaging (MRI) and X-ray computed tomography (CT) provide non-invasive techniques of investigating the human anatomy, thus significantly improving diagnosis and treatment of diseases. CT is well-suited for visualizing bone structures, including location and density. Bone features are essential for many advanced applications, \eg image-guided radiotherapy and reconstruction in electroencephalography (EEG). CT scans, however, carry a risk of causing cancer in the subject due to ionizing radiation. The additional risk of a subject developing fatal cancer from a CT scan is approximately 1 in 2000~\cite{CTriskStat}.
Notably, young individuals are more susceptible to radiation-induced diseases than adults~\cite{CTriskKids}.
MRI, on the other hand, does not expose the subject to ionizing radiation and is considered a safe procedure.
MRI further contrasts with CT in that it has excellent soft tissue contrast visualization but cannot image bone.
These facts motivate the development of methods in which a single image modality is used to synthesize the desired information, that could be provided by another imaging modality, with the end goal of making clinical processes more effective and circumventing unnecessary inconveniences for patients.

CT image synthesis was initially developed as a means for attenuation correction in positron emission tomography (PET)~\cite{RegSyn08}. Researchers still revisit registration-based synthesis methods due to its overall good performance and fast prediction times. Multi-atlas registration methods are capable of estimating the intensity distribution within bone with good approximation with prediction times within 5 minutes~\cite{RegSyn16,RegSyn17}.
Alongside the development of new registration-based methods, the development of deep learning~\cite{LeCun2015} has inspired a revolution of learning-based methods. The development of versatile models, originating from computer vision, has sparked an interest in deep learning techniques for medical image analysis.
Deep learning techniques for cross-modality medical image synthesis first outperformed existing learning based-methods such as k-nearest neighbor when using a relatively shallow convolutional neural network (CNN) to synthesize 3D PET images from an MRI~\cite{DataCompletion}.
Later, a 3D fully convolutional network (FCN) was proposed to synthesize a CT image from MRI~\cite{EstimateCTwithFCN}. By adopting up-pooling, the FCN preserves structural information such as neighboring pixel values. This FCN outperformed random forest and atlas-based registration methods.
Transfer learning also proved useful for synthesizing 2D CT slices from 2D MRI slices~\cite{sCTwithTransferLearning}. By initializing the model with the learned filters from a network trained on natural images, the final trained model performed better than atlas-based registration methods.

Conditional generative adversarial networks (cGAN) have further refined the estimation of CT images. In particular, an auto-context model (ACM) consisting of three separately trained cGANs was shown to estimate CT images more accurately than an FCN without a discriminator network~\cite{ContextAwareSynthesis}, with prediction times under four minutes.
Most recently, an FCN consisting of embedding blocks trained on 3D MRI and CT data, also called a deep embedding CNN (DECNN) has been proposed~\cite{EmbeddingSynthesis}. By using embedding blocks, the network is forced to output tentative CT images. This approach outperformed atlas-based registration methods and also deep learning methods such as a CNN and an FCN model.

We contribute to this area of research by presenting four methods capable of MRI to CT synthesis adapted from previous methods in the recent literature. We propose a novel framework in which multi-atlas registration synthesis serves as a prior to a deep neural network (DNN). We evaluate the synthesis results and give a general comparison of all methods. Finally, we show that segmentation of synthetic CT images is more accurate than learning to identify bone directly from the MRI.

%% file: tex/method.tex
An image is a function that maps $d$-dimensional points, from the set of points $\Omega$ in the image domain, to $m$-component intensity values $I_{\Omega}: \mathbb{R}^d \mapsto \mathbb{R}^m$. In image synthesis, we aim to estimate the function $S$ that best approximates the ground-truth CT image $I_\text{CT}$ from an MRI $I_\text{MR}$ of the same subject, \ie $I_\text{CT} \approx S(I_\text{MR})$.

\subsection{Synthesis Using Multi-Atlas Registration}
\label{sec:atlas}
We perform atlas-based image synthesis of a given subject's MRI $I_\text{MR}$ by registering a set of $N_\text{atlas}$ atlases, which consist of co-registered MR and CT image pairs along with gold-standard brain segmentation masks, to this image.
Drastic differences in the image field of view that result in cropped anatomy and missing correspondences between a subject's MRI and the various MR atlas images make intensity-based image registration challenging when optimizing metrics like normalized mutual information (NMI)~\cite{NMI}.
Instead, we perform a more robust surface-based atlas registration, where the brain surface $S_{\text{atlas},i}$ in each atlas is extracted from the segmentation mask and the brain surface $S_\text{MR}$ in $I_\text{MR}$ can reliably be found using standard skull stripping methods~\cite{BET}.
We then register $S_{\text{atlas},i}, i=1,\dots,N$ to  $S_\text{MR}$ using robust point matching~\cite{RPM}, and use the resulting transformations to warp $I_{\text{CT},i}$ to  $I_\text{MR}$. The transformed atlas CTs are then averaged to form the synthesized CT image.

\subsection{Deep Neural Network Regression}
\label{sec:dnn}
We treat image synthesis as a regression problem and fit a deep learning model to the distribution of CT images with MR images as a conditional distribution. Specifically, we train a DNN on MR images and backpropagate the error between the output of the DNN and the ground truth CT image. We adopt the U-Net architecture~\cite{U-Net}, as it ensures the preservation of high-frequency information while learning an internal representation of the input images. We propose to use a modified version of the U-Net, where we use 3 max pooling operations and 3 layers of convolutions instead of 2 prior to pooling (or up-sampling) for a total of 23 layers. We train the model with randomly sampled patches from the training set. Patch-wise training lowers memory restrictions while providing  the opportunity to augment the training data and maintaining variance in each mini-batch. The $L_2$ loss and image gradient difference loss (GDL) is used to penalize the error \cite{ContextAwareSynthesis}.
\begin{equation}
\mathcal{L}(I_\text{MR},I_\text{CT}) = \frac{1}{|\Omega|} \sum_{i \in \Omega} \left(\left(S(I_\text{MR})(i) - I_\text{CT}(i)\right)^2\right) + GDL(S(I_\text{MR}),I_\text{CT})
\end{equation}
where $\Omega$ is the set of all indices in $S(I_\text{MR})$, which is the same size as $I_\text{CT}$. GDL is defined as the sum of the absolute squared differences between the gradients of image $X$ and $Y$ in each dimension.
\begin{equation}
GDL(X,Y) = \frac{1}{|D||\Omega|} \sum_{d \in D} \sum_{i \in \Omega} \left|\nabla X_d (i) - \nabla Y_d (i)\right|^2
\end{equation}
where $D$ is the set of dimension of $X$ and $Y$. $\nabla X_d$ and $\nabla Y_d$ denote the image gradients with regards to dimension $d$.

A CT prediction is made by extracting overlapping patches from $I_\text{MR}$, forward passing all patches and reconstructing the output. The loss tends to increase with distance from the center of a patch, due to less contextual information in proximity of borders of the patch and padding. Therefore the reconstruction function applies a Gaussian distributed mask to weight center pixels more than pixels on the border of the patch.

\subsection{Conditional Generative Adversarial Network}
\label{sec:cGAN}
A conditional generative adversarial network (cGAN) includes two networks. The objective of the generator is to learn $S$ by approximating the ground truth CT images while producing CT images indistinguishable from the discriminator. The objective of the discriminator is to classify CT images as either synthesized or true CT images. In this work, the generator is the U-Net (Sec.~\ref{sec:dnn}). An adversarial loss term is added to the total generator loss $\mathcal{L_G}$. The classification output of the discriminator is penalized with a corresponding adversarial loss term
\begin{equation}
\begin{split}
\mathcal{L}_G(I_\text{MR},I_\text{CT}) = & \frac{1}{|\Omega|} \sum_{i \in \Omega} \left(
    \left(S(I_\text{MR})(i) - I_\text{CT}(i)\right)^2 \right)
\\ &+ GDL(S(I_\text{MR}),I_\text{CT})
\\ &+ 
\frac{1}{2}
\left(
    C(S(I_\text{MR})) - T_{\text{real}} 
\right)^2
\end{split}
\end{equation}
\begin{equation}
\begin{split}
\mathcal{L}_D(I_\text{MR},I_\text{CT}) = &
\frac{1}{2}
\left(
\left(
    C(I_\text{CT}) - T_{\text{real}} 
\right)^2 
+ 
\left(
    C(S(I_\text{MR})) - T_{\text{synthetic}}
\right)^2
\right)
\end{split}
\end{equation}
The adversarial loss terms define the cGAN as a least square cGAN which is more stable when training than a regular GAN with binary cross entropy loss~\cite{LSGAN}.
During training, a full forward pass is performed trough $G$ and $D$ and the errors are computed using $\mathcal{L}_G$, $\mathcal{L}_G$. The error defined by $\mathcal{L}_G$ in backpropagated through $G$ and the error defined by $\mathcal{L}_D$ in backpropagated through $D$. Weights of $G$ and $D$ are updated, in that order.

\subsection{Residual Learning with Atlas Prior}
\label{sec:boneweight}
\begin{figure}[t]
\centering
\includegraphics[width=\textwidth]{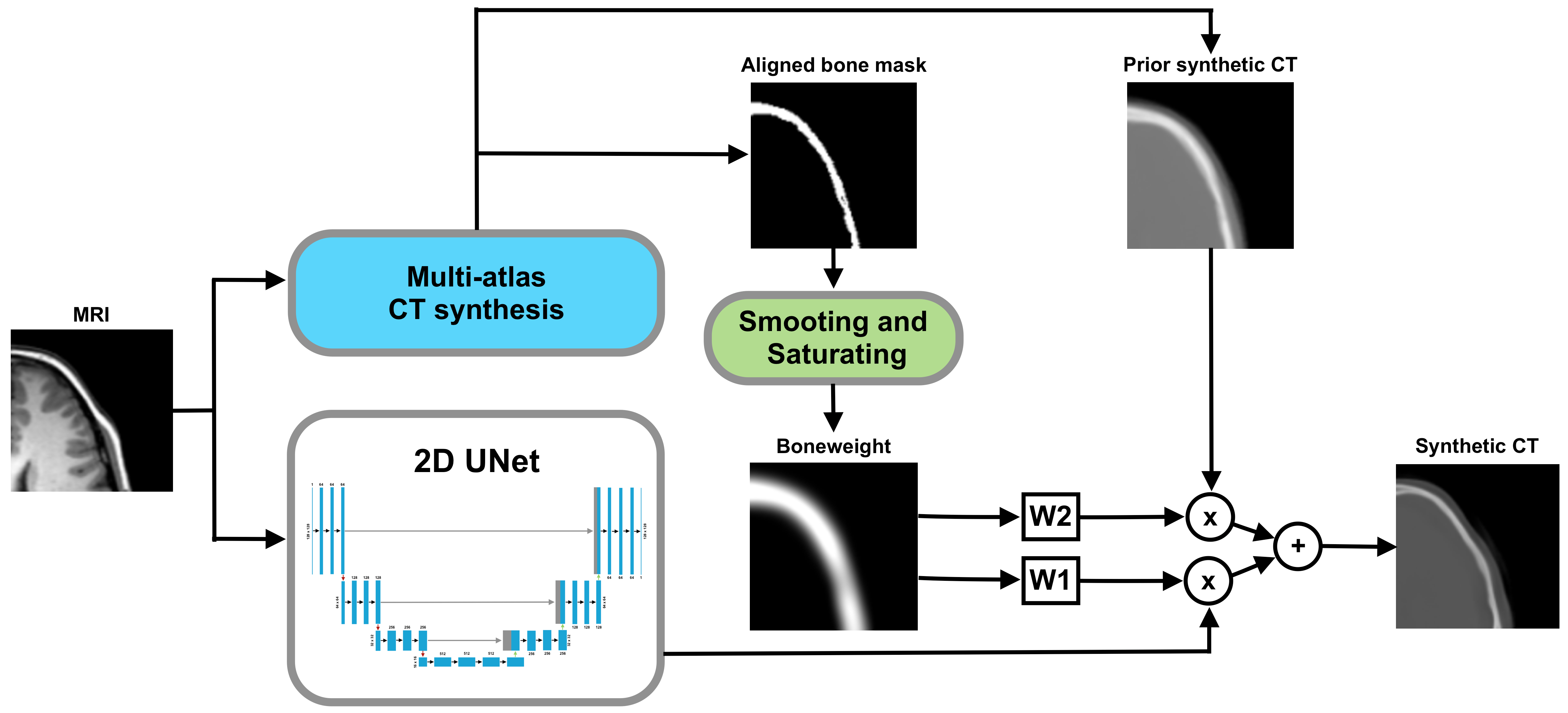}
\caption{Our proposed framework for MRI to CT image synthesis that incorporates prior information.}
\label{fig:framework}
\end{figure}

For most applications utilizing synthetic CT images, the primary area of interest is bone features such as bone location and density. The soft tissue is often less significant and can be rapidly estimated with atlas-based synthesis.
We propose a framework (Fig.~\ref{fig:framework}) capable of synthesizing a CT image $S(I_\text{MR})$ in three stages; i) synthesizing a prior synthetic CT image, $S_\text{atlas}(I_\text{MR})$, with atlas-based registration ii) computing the difference, $S_\text{DNN}(I_\text{MR})$, between $S(I_\text{MR})$ and $S_\text{atlas}(I_\text{MR})$ with a DNN and iii) combine of $S_\text{atlas}(I_\text{MR})$ and $S_\text{DNN}(I_\text{MR})$ with a weighted addition:
\begin{equation}
S(I_\text{MR}) = W_1 S_\text{DNN}(I_\text{MR}) + W_2 S_\text{atlas}(I_\text{MR})
\end{equation}
The prior synthetic CT is obtained by affine multi-atlas registration as described in section \ref{sec:atlas}. Additionally, the transformation is not only applied to the atlas CT images but also the reference bone masks, which are later used to define the weights $W_1$ and $W_2$.
The aligned bone masks are merged into a single mask and smoothed with a Gaussian distribution parameterized by the standard deviation $\sigma$. The smoothed bone mask, $W_\text{bone}$, is then saturated with the function:
\begin{equation}
W_\text{bone}(i) = \bigg\{\begin{array}{lr}
        1 & \;\;\text{if}\;\; I_\text{bone}(i) > t \\
        \frac{I_\text{bone}(i)}{t} & \;\;\text{if}\;\; I_\text{bone}(i) \leq t
        \end{array},\;\; \forall i \in \Omega_{I_\text{bone}}
\end{equation}
where $t$ is a threshold such that $0 \leq t \leq 1$ and $I_\text{bone}(i)$ is the intensity value at index $i$ from the set of indices, $\Omega_{I_\text{bone}}$, in $I_\text{bone}$. $W_\text{bone}$ provides an image that roughly locates bone. The weights are computed by the following linear relationship:
\begin{equation}
W = \alpha W_\text{bone} + \beta\
\end{equation}
where $\alpha$ and $\beta$ are to be chosen such that $0 \leq W_1 \leq 1$. These parameters allow tuning of how much to weigh $S_\text{DNN}(I_\text{MR})$ and how much to weigh $S_\text{atlas}(I_\text{MR})$ and can be altered at any stage of the training. By using $W_\text{bone}$ as a weight on the output of a DNN and computing the loss after adding the prior synthetic CT, we force the it to pay less attention to areas outside the head and inside the skull. Again, the DNN is chosen to be the U-Net (Sec.~\ref{sec:dnn}), which computes the function $S_\text{DNN}(I_\text{MR})$, and is trained in the same manner as the model in Sec.~\ref{sec:dnn}.

%% file: tex/experiments.tex
\subsection{Data}
The dataset consists of $N=30$ paired 3D T1-weighted MR and CT images $\{I_{\text{MR},i},I_{\text{CT},i}\}$ from different pediatric subjects $i=1,\dots,N$ acquired retrospectively by data mining the clinical image repository of the Washington University BJC Health System (St. Louis, MO, USA) within the Pediatric Head Modeling project~\cite{PedHead}. Institutional Review Boards at both project sites (Philips-Electrical Geodesics, Inc and the University of Arkansas for Medical Sciences) approved all research retrospective protocols involving human subjects~\cite{DataProt}. The age range of the subjects is six months old to 16 years and five months old. For atlas-based synthesis, we use $N_\text{atlas}=5$ co-registered MR and CT image volumes with tissue masks, consisting of 3 adult and 2 pediatric subjects (\url{ www.egi.com}). 

We spatially normalized all images to a common template space in two phases: 
\begin{inparaenum}[(\itshape i\upshape)]
\item  inter-subject affine registration of all MR images to the MNI Colin 27 brain reference space using NMI~\cite{NMI}, and 
\item rigid intra-subject registration of all $\{I_{\text{MR},i},I_{\text{CT},i}\}$ pairs using NMI.
\end{inparaenum}
All images were resliced to template space with isotropic $1\text{mm}^3$ resolution.
We normalized the MR image intensity values by subtracting the mean and dividing by the standard deviation of the intensity values from within brain tissue (we used the Colin 27 brain mask to roughly estimate the brain volume in the spatially normalized images). 

\subsection{Similarity Metrics}
We evaluated the quality of the synthesized CT images by the quality of tissue in the whole head and, since we are most interested in the bone, we evaluated the quality of synthesis specifically in the proximity of bone. At prediction time the synthesized CT image is multiplied with a head mask and a smoothed bone mask to extract the areas of interest.
Given $X$ and $Y$ are the predicted and ground truth images respectively, we evaluate synthesis using the following similarity metrics:
\begin{inparaenum}[(\itshape i\upshape)]
\item peak signal-to-noise ratio (PSNR),
\begin{equation}
\text{PSNR}(X,Y) 
= 20 \log_{10}\left(\frac{v_{max}}{\sqrt{\frac{1}{|\Omega|}\sum_{i \in \Omega}(Y(i)-X(i))^2}}\right),
\end{equation}
where $v_{max} = 4096$ as CT images are normalized to the intensity range $[0,4095]$;
\item mean structural similarity (MSSIM),
\begin{equation}
\text{MSSIM}(X,Y) 
= \frac{1}{|\Omega|} \sum_{i \in \Omega} 
\frac{(2\mu_{X} \mu_{Y} + c_1) (2\sigma_{XY} + c_2)}
{(\mu_X^2 + \mu_Y^2 + c_1) (\sigma_X^2 + \sigma_Y^2 + c_2)},
\end{equation}
where $\mu_{I}$ and $\sigma_I$ are the mean and the standard deviation, respectively, of image $I$ and constants $c_1 = (0.01\cdot(2^{12}-2))^2$ and $c_2 = (0.03\cdot(2^{12}-2))^2$; 
\item mean average error (MAE), 
\begin{equation}
\text{MAE}(X,Y)
= \frac{1}{|\Omega|} \sum_{i \in \Omega} |Y(i) - X(i)|;
\end{equation}
\item Pearson cross-correlation (PCC),
\begin{equation}
\text{PCC}(X,Y)
= \frac{\sum_{i \in \Omega}(X(i) - \mu_X)(Y(i) - \mu_Y)}{\sqrt{\sum_{i \in \Omega}(X(i) - \mu_X)^2} \sqrt{\sum_{i \in \Omega}(Y(i) - \mu_Y)^2}};
\end{equation}
and, as an additional method of evaluating the quality of synthesized CT images, we segment both the ground truth CT image and the synthesized CT image into three classes: air, bone, and soft-tissue. We perform the segmentation with the k-means clustering of intensity values distributed in 1024 bins. Synthesized CT images obtained with deep learning algorithms are not guaranteed to be Hounsfield scaled, and the threshold segmentation might not apply. The masks for bone are extracted, and we evaluate with the DICE overlap metric,
\begin{equation}
\text{DICE}(B_X,B_Y)
= \frac{2 TP}{2TP + FP + FN}
\end{equation}
Where $B_X$ and $B_Y$ are the bone mask for the prediction image and the ground truth image respectively. $TP$, $FP$ and $FN$ are true positives, false positives and false negatives respectively. This metric quantifies how well the bone can be segmented in the synthetic CT images.
Segmenting the ground truth CT images with the mentioned k-means algorithm and computing the DICE overlap with ground truth bone mask yields a mean (standard deviation) and median of $0.91 (0.03)$ and $0.92$ respectively. This will be the baseline by which we compare our results where a DICE overlap of $0.91$ is the ideal score.
\end{inparaenum}

\subsection{Experiments}
\begin{figure}[t]
\includegraphics[width=\textwidth]{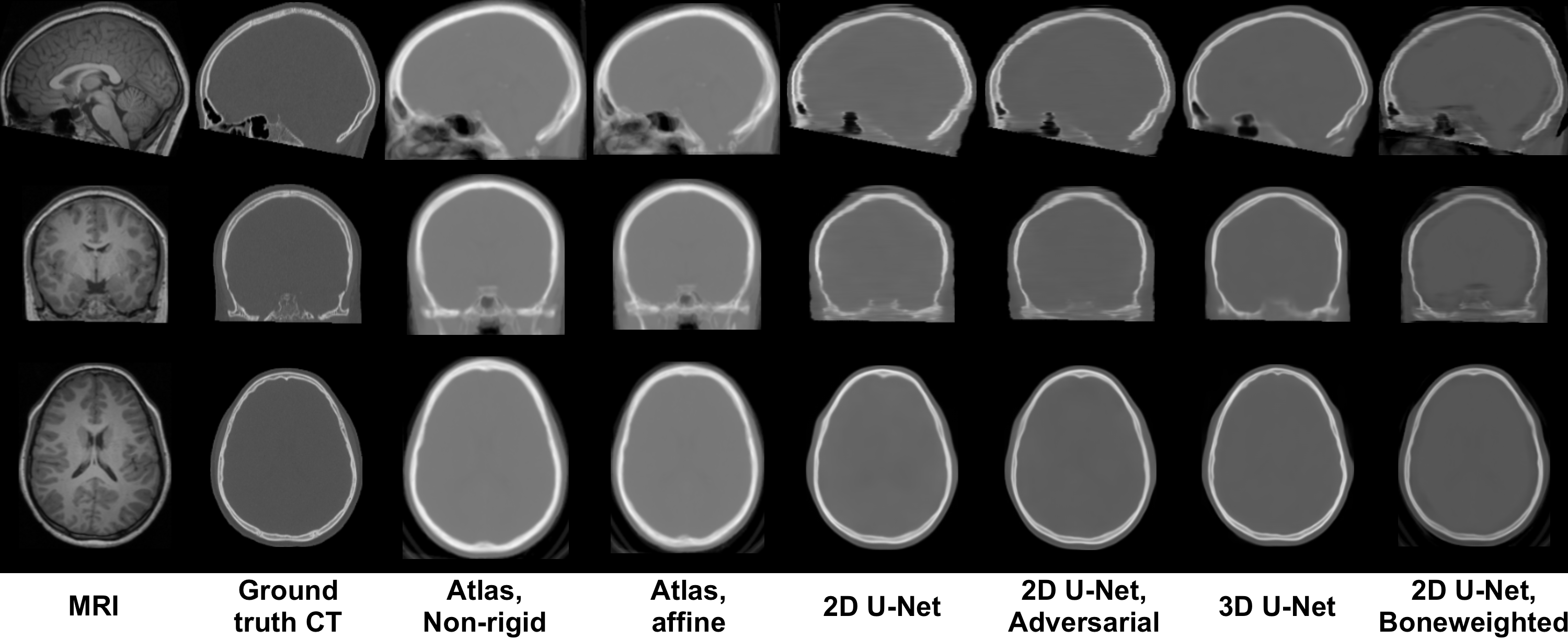}
\caption{Same subject center slices from MRI, ground truth CT and synthetic CT from all methods.}
\label{fig:images}
\end{figure}

We trained and tested the deep learning models using 6-fold cross validation with 25 images in the training set and five images in the test set. Each epoch, 3200 randomly sampled patches were extracted from the training set and used for training. We compared models using both 2D patches and 3D patches, and used patch sizes of $128 \times 128$ and $48 \times 48 \times 24$, respectively. We performed data augmentation by randomly flipping the patches in the $x$ and $y$. We trained the models for 650 epochs with the ADAM optimizer and fixed learning rate of $0.0001$. For the registration-based multi-atlas image synthesis (Sec.~\ref{sec:atlas}), we registered the 5 atlas images to each of the 30 test MR images using both affine and non-rigid transformations. For the non-rigid transformations, we used free-form deformation (FFD)~\cite{FFD} with 30mm B-spline control point spacing.
Synthetic CT images from all methods are displayed in Fig.~\ref{fig:images}. 
Tables~\ref{res:head} and~\ref{res:bone} show synthesis evaluation results for the head and bone areas, respectively, and Table~\ref{res:seg} shows bone segmentation results.
We also compared segmentation of the bone directly in the MRI by training a U-Net with weighted cross-entropy as the loss function, for 160 epochs, with the same parameters as described above (we label this 2D U-Net, direct MRI segmentation).

\begin{table}[t]
\caption{Results of MRI to CT synthesis with all methods evaluated with PSNR, MAE, MSSIM and PCC on the head masked synthetic CT.}
\centering
\resizebox{\textwidth}{!}{%
\begin{tabular}{lcccccccc}
\hline
\multirow{2}{*}{\begin{tabular}[c]{@{}l@{}}Similarity Metric\end{tabular}} & \multicolumn{2}{c}{PSNR} & \multicolumn{2}{c}{MAE} & \multicolumn{2}{c}{SSIM} & \multicolumn{2}{c}{PCC} \\ \cline{2-9} 
 & Mean (std) & Median & Mean (std) & Median & Mean (std) & Median & Mean (std) & Median \\ \hline
\begin{tabular}[c]{@{}l@{}}Atlas, non-rigid\end{tabular} & 25.85 (0.95) & 25.82 & 71.66 (13.76) & 71.45 & 0.9 (0.03) & 0.89 & 0.94 (0.02) & 0.94 \\
\begin{tabular}[c]{@{}l@{}}Atlas, affine\end{tabular} & 25.97 (1.01) & 26 & 71.61 (14.79) & 69.95 & 0.89 (0.03) & 0.9 & 0.94 (0.02) & 0.94 \\
2D U-Net & 28.01 (2.04) & 27.99 & 44.58 (16.64) & 42.66 & 0.91 (0.03) & 0.91 & 0.951 (0.023) & 0.96 \\
\begin{tabular}[c]{@{}l@{}}2D U-Net, adversarial\end{tabular} & 27.93 (1.94) & 27.92 & 44.75 (16.34) & 42.77 & 0.91 (0.03) & 0.91 & 0.95 (0.02) & 0.95 \\
3D U-Net & 27.9 (1.96) & 27.84 & 44.97 (15.38) & 42.2 & 0.91 (0.03) & 0.91 & 0.95 (0.02) & 0.95 \\
\begin{tabular}[c]{@{}l@{}}2D U-Net, boneweighted\end{tabular} & 27.38 (1.84) & 27.54 & 56.04 (16.86) & 54 & 0.9 (0.03) & 0.91 & 0.95 (0.02) & 0.95 \\ \hline
\end{tabular}%
}
\label{res:head}
\end{table}

\begin{table}[t]
\caption{Results of MRI to CT synthesis with all methods evaluated with PSNR, MAE, MSSIM and PCC on the bone masked synthetic CT.}
\centering
\resizebox{\textwidth}{!}{%
\begin{tabular}{lcccccccc}
\hline
\multirow{2}{*}{\begin{tabular}[c]{@{}l@{}}Similarity Metric\end{tabular}} & \multicolumn{2}{c}{PSNR} & \multicolumn{2}{c}{MAE} & \multicolumn{2}{c}{SSIM} & \multicolumn{2}{c}{PCC} \\ \cline{2-9} 
 & Mean (std) & Median & Mean (std) & Median & Mean (std) & Median & Mean (std) & Median \\ \hline
\begin{tabular}[c]{@{}l@{}}Atlas, non-rigid\end{tabular} & 38.82 (1.67) & 38.78 & 16.35 (4.11) & 16.08 & 0.93 (0.01) & 0.93 & 0.9 (0.03) & 0.9 \\ 
\begin{tabular}[c]{@{}l@{}}Atlas, affine\end{tabular} & 38.87 (1.71) & 38.89 & 16.29 (4.23) & 15.88 & 0.93 (0.02) & 0.93 & 0.9 (0.03) & 0.9 \\ 
2D U-Net & 41.61 (3.02) & 41.52 & 9.09 (4.8) & 8.25 & 0.97 (0.02) & 0.97 & 0.94 (0.03) & 0.94 \\ 
\begin{tabular}[c]{@{}l@{}}2D U-Net, adversarial\end{tabular} & 41.59 (2.88) & 41.41 & 9.03 (4.5) & 8.31 & 0.97 (0.02) & 0.97 & 0.94 (0.02) & 0.94 \\ 
3D U-Net & 41.51 (2.98) & 41.53 & 9.21 (4.64) & 8.12 & 0.97 (0.02) & 0.97 & 0.94 (0.02) & 0.94 \\ 
\begin{tabular}[c]{@{}l@{}}2D U-Net, boneweighted\end{tabular} & 41.27 (2.77) & 41.19 & 10.06 (4.53) & 9.25 & 0.96 (0.02) & 0.96 & 0.93 (0.03) & 0.94 \\ \hline
\end{tabular}%
}
\label{res:bone}
\end{table}

\begin{table}[t]
\caption{Dice overlap scores of segmenting synthetic CT image from all methods with k-means clustering compared to direct MRI segmentation.}
\centering
\resizebox{\textwidth-3.5cm}{!}{%
\begin{tabular}{lcc}
\hline
\multirow{2}{*}{Similarity metric} & \multicolumn{2}{c}{Dice} \\ \cline{2-3} 
 & Mean (std) & Median \\ \hline
\begin{tabular}[c]{@{}l@{}}Atlas, non-rigid\end{tabular} & 0.56 (0.08) & 0.56 \\ 
\begin{tabular}[c]{@{}l@{}}Atlas, affine\end{tabular} & 0.55 (0.08) & 0.53 \\ 
2D U-Net & 0.64 (0.11) & 0.63 \\ 
\begin{tabular}[c]{@{}l@{}}2D U-Net, adversarial\end{tabular} & 0.65 (0.1) & 0.64 \\ 
3D U-Net & 0.64 (0.1) & 0,63 \\ 
\begin{tabular}[c]{@{}l@{}}2D U-Net, boneweighted\end{tabular} & 0.63 (0.11) & 0.63 \\ 
\begin{tabular}[c]{@{}l@{}}2D U-Net, direct MRI segmentation\end{tabular} & 0.63 (0.1) & 0.63 \\ 
\hline
\end{tabular}
}
\label{res:seg}
\end{table}

%% file: tex/discussion.tex
We have performed an extensive evaluation of several methods capable of MR to CT image synthesis with visually good results.
The multi-atlas image synthesis methods yielded perceptually less satisfying synthetic images than the deep learning models, as seen in Fig.~\ref{fig:images}, which was also reflected in across all metrics including segmentation performance.
We demonstrated that 3D deep learning models are not better at synthesizing CT images than a corresponding 2D model. The 2D model achieved a mean PSNR of $28.01 \pm 2.04\text{ dB}$ evaluated on the head and $41.69 \pm 3.02\text{ dB}$ on the area around the bone. The 3D model achieved a PSNR of $27.9 \pm 1.96\text{ dB}$ evaluated on the head and $41.51 \pm 2.98\text{ dB}$ on the area in proximity of bone. The 3D model took 54 hours to train while the 2D model took 16 hours to train. Predicting an image took 10 seconds with the 2D model and 50-65 seconds with the 3D model. Synthetic CT images produced by the 3D model looks visually better in the third axis.
The 2D U-Net model and the adversarial 2D U-Net model resulted in very similar metric scores and also visually similar images, but the adversarial 2D U-Net model took 30 hours to train.
Adopting a 2D U-Net for synthesizing CT was the best method concerning memory consumption, training time and prediction time. The method achieved better or equally as high metric scores as all the other methods and produces visually outstanding CT images, and this suggests that the choice of deep neural network architecture plays a crucial role for performance. 

Furthermore, we demonstrated that segmenting synthetic CT images produced by a 2D U-Net resulted in a higher DICE overlap than training a 2D U-Net to segment bone directly from MR images. Still, the DICE overlap score was low (under $0.65$ of possible $0.91$). MRI bone segmentation is an ill-posed and challenging problem, and there is room for significant improvement in this area of research.

We presented a novel framework for synthesizing CT images. These images are perceptually less satisfying than the other methods, as borders from the atlas are visible. In future work, we will improve on this novel method using atlas priors to deep learning models such that we will be able to train a model with significantly fewer parameters that are specifically tailored to synthesize bone features and will enable better bone segmentation. 

%% file: tex/acknowledgements.tex
\subsubsection*{Acknowledgements}
This work was supported by the National Institute of Health (NIH) National Institute of Neurological Disorders and Stroke (NINDS) R44 NS093889.
Collection of the data partially used in this work was supported by NIH NINDS R43 NS67726, and the National Institute of Mental Health R44 MH106421.
Additionally, this work was supported by grants from Knud Højgårds Foundation, The Lundbeck Foundation, The Oticon Foundation and Family Hede Nielsen Foundation.

%% file: main.bbl
\begin{thebibliography}{10}
\providecommand{\url}[1]{\texttt{#1}}
\providecommand{\urlprefix}{URL }
\providecommand{\doi}[1]{https://doi.org/#1}

\bibitem{RegSyn17}
Burgos, N., Guerreiro, F., McClelland, J., Presles, B., Modat, M., Nill, S.,
  Dearnaley, D., deSouza, N., Oelfke, U., Knopf, A.C., Ourselin, S., Cardoso,
  M.J.: Iterative framework for the joint segmentation and ct synthesis of mr
  images: Application to mri-only radiotherapy treatment planning. Physics in
  Medicine and Biology  \textbf{62},  4237--4253 (03 2017)

\bibitem{sCTwithTransferLearning}
Han, X.: Mr-based synthetic ct generation using a deep convolutional neural
  network method. Medical Physics  \textbf{44},  1408--1419 (02 2017)

\bibitem{RegSyn08}
Hofmann, M., Steinke, F., Scheel, V., Charpiat, G., Farquhar, J., Aschoff, P.,
  Brady, M., Sch{\"o}lkopf, B., Pichler, B.J.: Mri-based attenuation correction
  for pet/mri: a novel approach combining pattern recognition and atlas
  registration. Journal of Nuclear Medicine  \textbf{49 11},  1875--83 (2008)

\bibitem{CTriskKids}
Kutanzi, K.R., Lumen, A.A., Koturbash, I., Miousse, I.R.: Pediatric exposures
  to ionizing radiation: Carcinogenic considerations. In: International journal
  of environmental research and public health (2016)

\bibitem{LeCun2015}
LeCun, Y., Bengio, Y., Hinton, G.: {Deep learning}. Nature  \textbf{521}(7553),
   436--444 (2015)

\bibitem{DataCompletion}
Li, R., Zhang, W., Suk, H., Wang, L., Li, J., Shen, D., Ji, S.: Deep learning
  based imaging data completion for improved brain disease diagnosis. Medical
  image computing and computer-assisted intervention : MICCAI  \textbf{17}(Pt
  3),  305--312 (2014)

\bibitem{LSGAN}
Mao, X., Li, Q., Xie, H., Lau, R.Y.K., Wang, Z., Smolley, S.P.: Least squares
  generative adversarial networks. In: 2017 IEEE International Conference on
  Computer Vision (ICCV). pp. 2813--2821 (2017)

\bibitem{EstimateCTwithFCN}
Nie, D., Cao, X., Gao, Y., Wang, L., Shen, D.: Estimating ct image from mri
  data using 3d fully convolutional networks. In: Deep Learning and Data
  Labeling for Medical Applications. pp. 170--178. Springer International
  Publishing (2016)

\bibitem{ContextAwareSynthesis}
Nie, D., Trullo, R., Lian, J., Petitjean, C., Ruan, S., Wang, Q., Shen, D.:
  Medical image synthesis with context-aware generative adversarial networks.
  In: Medical Image Computing and Computer-Assisted Intervention. pp. 417--425.
  Springer International Publishing (2017)

\bibitem{CTriskStat}
NIH: Computed tomography ({CT}) scans and cancer (2013),
  \url{https://www.cancer.gov/about-cancer/diagnosis-staging/ct-scans-fact-sheet#r2}
  (visited 2018-05-08)

\bibitem{RPM}
Rangarajan, A., Chui, H., Mjolsness, E., Pappu, S., Davachi, L., Goldman-Rakic,
  P., Duncan, J.: {A robust point-matching algorithm for autoradiograph
  alignment}. Medical Image Analysis  \textbf{1}(4),  379--398 (1997)

\bibitem{U-Net}
Ronneberger, O., Fischer, P., Brox, T.: U-net: Convolutional networks for
  biomedical image segmentation. In: International Conference on Medical image
  computing and computer-assisted intervention. pp. 234--241. Springer (2015)

\bibitem{FFD}
Rueckert, D., Sonoda, L., Hayes, C., Hill, D., Leach, M., Hawkes, D.: {Nonrigid
  registration using free-form deformations: application to breast MR images}.
  IEEE Transactions on Medical Imaging  \textbf{18}(8),  712--721 (1999)

\bibitem{BET}
Smith, S.M.: {Fast robust automated brain extraction}. Human Brain Mapping
  \textbf{17}(3),  143--155 (2002)

\bibitem{DataProt}
Song, J., Morgan, K., Sergei, T., Li, K., Davey, C., Govyadinov, P.:
  Anatomically accurate head models and their derivatives for dense array eeg
  source localization. Functional Neurology, Rehabilitation, and Ergonomics
  \textbf{3},  275--294 (01 2013)

\bibitem{NMI}
Studholme, C., Hill, D.L.G., Hawkes, D.J.: {An overlap invariant entropy
  measure of {\{}3D{\}} medical image alignment}. Pattern Recognition
  \textbf{32}(1),  71--86 (1999)

\bibitem{RegSyn16}
Torrado-Carvajal, A., Herraiz, J.L., Alcain, E., Montemayor, A.S.,
  Garcia-Cañamaque, L., Hernandez-Tamames, J.A., Rozenholc, Y., Malpica, N.:
  Fast patch-based pseudo-ct synthesis from t1-weighted mr images for pet/mr
  attenuation correction in brain studies. Journal of Nuclear Medicine
  \textbf{57}(1),  136--143 (2016)

\bibitem{PedHead}
Turovets, S.: https://www.pedeheadmod.net/ (2018),
  \url{https://www.pedeheadmod.net/} (visited 2018-06-21)

\bibitem{EmbeddingSynthesis}
Xiang, L., Wang, Q., Nie, D., Zhang, L., Jin, X., Qiao, Y., Shen, D.: Deep
  embedding convolutional neural network for synthesizing ct image from
  t1-weighted mr image. Medical Image Analysis  \textbf{47},  31 -- 44 (2018)

\end{thebibliography}
